\documentclass[prd,floats,floatfix,nofootinbib]{revtex4}
\usepackage[dvips]{graphicx}
\usepackage{graphics}
\usepackage{dcolumn}
\usepackage{amssymb}
\usepackage{bm}
\usepackage{amsmath}
\usepackage{color}
\usepackage{epstopdf}

\def\spose#1{\hbox to 0pt{#1\hss}}

\def\lta{\mathrel{\spose{\lower 3pt\hbox{$\mathchar"218$}}
     \raise 2.0pt\hbox{$\mathchar"13C$}}}
\def\gta{\mathrel{\spose{\lower 3pt\hbox{$\mathchar"218$}}
     \raise 2.0pt\hbox{$\mathchar"13E$}}}
\newcommand{\be}{\begin{equation}}
\newcommand{\en}{\end{equation}}
\newcommand{\bea}{\begin{eqnarray}}
\newcommand{\ena}{\end{eqnarray}}

\begin{document}

\title{Curvature Late-Time Acceleration in an Eternal Universe}

\author{Rodrigo Maier\footnote{rodrigo.maier@uerj.br} and
Ivano Dami\~ao Soares\footnote{ivano@cbpf.br}
\vspace{0.5cm}}

\affiliation{$^*$ Departamento de F\'isica Te\'orica, Instituto de F\'isica, Universidade do Estado do Rio de Janeiro,\\
Rua S\~ao Francisco Xavier 524, Maracan\~a,\\
CEP20550-900, Rio de Janeiro, Brazil\\
\\
$^\dagger$Centro Brasileiro de Pesquisas F\'{\i}sicas -- CBPF, \\ Rua Dr. Xavier Sigaud, 150, Urca,
CEP22290-180, Rio de Janeiro, Brazil}


\date{\today}

\begin{abstract}
We construct a FLRW universe considering an anisotropic
scaling between space and time at extremely high and low energies only.
In this context, Friedmann equations contain an additional term arising from spatial curvature which implements nonsingular bounces in the early Universe. The matter content of the model is a nonrelativistic pressureless perfect fluid and radiation.
By breaking covariance diffeomorphism also at extreme large scales,
an additional term furnishes late-time acceleration due to spatial curvature so that a cosmological
constant is not needed. In order to probe the final fate of the universe we also introduce a lower order curvature
term which dominates in deep IR. Given the observational parameters we obtain a concrete model in eternal recurrence
in which the end of late-time acceleration takes place at a redshift $z \simeq -0.14$ and the
universe recollapses at $z\simeq - 0.32$.
\end{abstract}
\maketitle
\section{Introduction}

Although General Relativity is the most successful theory that currently describes
gravitation, it presents some intrinsic pathologies when one tries to construct a
cosmological model from a proper theory of gravitation. In cosmology for instance, the $\Lambda{\it CDM}$
model gives us important predictions concerning
the evolution of the Universe and its current state\cite{mukhanov}. However, let us
assume that the initial conditions of our Universe were fixed when the early
Universe emerged from the semi-Planckian regime and started its classical
expansion. Evolving back such initial conditions using the classical Einstein field
equations it results that our Universe is driven towards an initial singularity in high UV
where the classical regime is no longer valid\cite{wald}.
On the other hand, since $1998$ observational data\cite{riess} are giving support to the
highly unexpected assumption that our Universe is currently in a state of accelerated expansion.
In order to explain this late-time acceleration phase cosmologists have been considering
the existence of a new field -- known as dark energy -- that violates the strong energy condition
in deep IR.
Although the cosmological constant seems to be the simplest and most
appealing candidate for dark energy, it poses a huge problem to quantum
field theory on how to accommodate its observed value
with vacuum energy calculations\cite{ccp}.
\par
During recent decades theories of gravitation have been considered in
order to solve such problems, by properly modifying General Relativity
either in the high UV, or in the deep IR limit.
In this context, for instance, bouncing models may circumvent the problem of the
cosmological singularity, solve the flatness and horizon problems\cite{nelson0} and
reproduce the power spectrum of primordial cosmological perturbations
inferred by observations\cite{nelson}. Different candidates for dark energy
may also be proposed in the realm of modified theories of gravitation\cite{lta}.
\par
In an attempt to construct a quantum theory of gravitation, P. Ho\v{r}ava proposed a
modified gravity theory by considering a Lifshitz-type anisotropic scaling between space and time
at high energies\cite{Horava}. In this context it has been shown\cite{cal,brand,maier,maier1} that higher order spatial
curvature terms can lead to regular bounces in the early universe and also to a complex cosmological
dynamics. This bounce feature is due to the presence of positive powers of the spatial curvature -- in the potential of the
theory -- that engender nontrivial modifications of the dynamics in the high UV limit.
\par
In this paper we intend to discuss an extension of such Ho\v{r}ava-Lifshitz bouncing models.
This extension corresponds to also consider negative powers of the spatial curvature in
the potential so that corrections in the deep IR limit are also obtained. From the theoretical point of view,
we remark that such assumption characterizes a departure from Ho\v{r}ava-Lifshitz gravity
which we intend to better address in a future analysis. Apart from such technical aspects,
in this paper we perform a first phenomenological investigation of these features.
%
%
%

\section{Field Equations and the model}
In the case of a 4-dimensional (1 + 3) spacetime, our basic assumption is that a preferred foliation of
spacetime is a priori imposed. Therefore it is natural to work with the
Arnowitt-Deser-Misner (ADM) decomposition of spacetime\cite{adm}
\begin{eqnarray}
\label{ds2}
ds^2=N^2dt^2-{^{(3)}g_{ij}}(N^i dt+dx^i)(N^j dt+dx^j)
\end{eqnarray}
where $N = N(t,x_i)$ is the lapse function, $N_i = N_i(t, x_i)$ is the shift and ${^{(3)}g_{ij}}(t, x_i)$
is the spatial geometry. Although the final action of the theory is not expected to be invariant
under diffeomorphisms as in General Relativity, an invariant foliation
preserving diffeomorphisms can be assumed. This is achieved if the action is invariant
under time reparametrization together with time-dependent spatial
diffeomorphisms, namely,
\begin{eqnarray}
\label{fpd}
t\rightarrow \tilde{t}(t),~x^i \rightarrow \tilde{x}^i(x^i,t).
\end{eqnarray}
It follows that the only covariant object under spatial diffeomorphisms that contains
one time derivative of the spatial metric is the extrinsic curvature $K_{ij}$, defined as
\begin{eqnarray}
\label{eqk}
K_{ij}=\frac{1}{2N}\Big(\frac{\partial{g}_{ij}}{\partial t}-\nabla_i N_j-\nabla_j N_i\Big)
\end{eqnarray}
where $\nabla_i$ is the covariant derivative relative to the spatial metric ${^{(3)}g_{ij}}$. Thus, to construct
a general theory which is of second order in time derivatives, one needs to consider
the quadratic terms $K_{ij}K^{ij}$ and $K^2$, where $K$ is the trace of $K_{ij}$. According to the previous assumptions,
a preferred foliation provides enough gauge freedom that allows us to fix $N_i=0$. In order to simplify our analysis,
we then propose the following action
\begin{eqnarray}
\label{eq1}
S = \int N \sqrt{{^{(3)}g}}~[K_{ij} K^{ij} - \lambda K^2 - {^{(3)}R} - U_{HL}(^{(3)}g_{ij},N) - 2\kappa^2{\cal L}_m] d^3x~ dt
\end{eqnarray}
where ${^{(3)}g}$ is the determinant of the spatial metric ${^{(3)}g}_{ij}$, ${^{(3)}R}$ is the spatial Ricci scalar and $\kappa^2$ is Einstein's constant. ${\cal L}_m$ is the lagrangian for the matter content of the model and $\lambda$
is a constant which corresponds to a dimensionless running coupling\cite{Horava}. Since in General Relativity the term $K_{ij} K^{ij} - K^2$
is invariant under 4-dim diffeomorphisms we expect to recover the classical regime as $\lambda \rightarrow 1$. For this reason
in the remaining of the paper we will consider $\lambda=1$.
\par
In general the potential $U_{HL}(^{(3)}g_{ij}, N)$ can depend on the spatial
metric and the lapse function due to the symmetry of the theory. It is clear that there are several invariant terms that can be
included in $U_{HL}$ -- particular choices often result in different versions of the $HL$ gravity.
However, in the following we intend to extend the HL scenario by imposing that $U_{HL}$ is a smooth function of ${^{(3)}R}$ only, namely, $U_{HL}=U_{HL}(^{(3)}R)$.
Finally ${\cal{L}}_m$ is the Lagrangean density of the matter content of the model, which we
take as noninteracting perfect fluids.
\par
As the fundamental symmetry assumed provides enough gauge freedom to choose $N=N(t)$ and $N_i=0$, we consider the case of a Friedmann-Lema\^itre-Robertson-Walker (FLRW) universe that in comoving
coordinates $x^{i}=(r,\theta,\phi)$ is expressed as
\begin{eqnarray}
\label{metric}
ds^2=N^2 dt^2+{^{(3)}g_{ij}}dx^{i} dx^{j}
\end{eqnarray}
where $t$ is the cosmological time,
\begin{eqnarray}
\label{metric1}
{^{(3)}g_{ij}}= - a^2(t) ~{\rm {diag}}~ \Big (\frac{1}{1-kr^2}, r^2,r^2 \sin^2{\theta} \Big),
\end{eqnarray}
$a(t)$ is the scale factor of the model and the parameter $k$ is proportional to the curvature of the 3-dim
spatial sections $t={\rm const}$.
The associated extrinsic curvature is given by
\begin{eqnarray}
\label{Excurv1}
{K_{ij}}= \frac{1}{2N}~ {^{(3)}{\dot{g}}_{ij}}= -\frac{a \dot{a}}{N} ~{\rm {diag}}~ \Big (\frac{1}{1-kr^2}, r^2,r^2 \sin^2{\theta} \Big).
\end{eqnarray}
%
%
%
%
%
\par
From (\ref{eq1}) we obtain the action of the model
\begin{eqnarray}
\label{Action2}
S= {V_0} \int  N a^3 \Big(  K_{ij}K^{ij}-K^2 -{^{(3)}}R  - U_{HL} - 2\kappa^2{\cal L}_m\Big) dt,
\end{eqnarray}
where $V_0$ is the spatial volume integral
\begin{eqnarray}
\nonumber
V_0=\int \frac{r^2 \sin  \theta}{\sqrt{1-kr^2}}~~dr d\theta d\phi .
\end{eqnarray}
The matter content of the model we take noninteracting dust and radiation,
namely,
\begin{eqnarray}
\label{lm}
{\cal L}_m=\rho_m+\rho_r,
\end{eqnarray}
where
\begin{eqnarray}
\label{mrn}
\rho_m = \rho_{0m}\Big(\frac{a_{0}}{a}\Big)^3, \rho_r = \rho_{0r}\Big(\frac{a_{0}}{a}\Big)^4.
\end{eqnarray}
The ``$0$'' subscript denotes the present epoch of our Universe. In order to simplify our analysis we will fix the natural normalization $a_0=1$.
From eqs. (\ref{Action2}) we then obtain
\begin{eqnarray}
\label{L1}
S = V_0 \int {\cal L} ~dt
\end{eqnarray}
where
\begin{eqnarray}
{\cal L}=  -\frac{6}{N} ~a {\dot a}^2 - N a^3 \Big [{^{(3)}}R +U_{HL}+2\kappa^2( \rho_m+\rho_r)  \Big].
\end{eqnarray}
\par By defining the canonical momentum
\begin{eqnarray}
\label{momentum}
p_a= \frac{\partial {\cal{L}}}{\partial {\dot{a}}}= - \frac{12 a {\dot{a}}}{N},
\end{eqnarray}
the total action can be expressed as
\begin{eqnarray}
\label{H}
S = V_0 \int \Big( {\dot{a}} p_a - N {\cal{H}} \Big)dt,
\end{eqnarray}
so that $\delta S/ \delta N =0$ results in the first integral of motion, the conserved
Hamiltonian constraint
\begin{eqnarray}
\label{Hamilton}
{\cal{H}}= -\frac{p_{a}^2}{24 a} + V(a) =0,
\end{eqnarray}
where
\begin{eqnarray}
\label{Hamilton1}
{V}(a)= \frac{2\kappa^2 \rho_{0r}}{a}+ 2\kappa^2 \rho_{0m} -6 k a +a^3 U_{HL}.
\end{eqnarray}
\par As mentioned already there are
several invariant terms that can be included in $U_{HL}$. For the purposes of the
present paper we will assume that $U_{HL}$ is a smooth
function of ${^{(3)}R}$ only, namely, $U_{HL}=U(^{(3)}R)$.
If we take into account that ${^{(3)}}R =-6k/ a^{2}$,
we see that the assumption of positive powers of $^{(3)}R$ in $U_{HL}$ leads to
$UV$ corrections -- this turns to be the case of Ho\v{r}ava-Lifshitz gravity\cite{cal, brand} for
bouncing cosmologies -- while negative powers of $^{(3)}R$ lead to IR corrections.
In the context of nonsingular cosmology, the core of this work
is to extend the scenario of bouncing models in HL gravity by considering terms in the potential such that
corrections in the deep IR are also obtained. For that matter, we will also assume
negative powers of $^{(3)}R$ in the potential $U$ which may be connected to a late-time acceleration regime.
%
\par
In a first analysis, we will assume here that the leading terms in the potential are
\begin{eqnarray}
\label{UU}
\label{u}
U(^{(3)}R)=\alpha_{0}\frac{1}{^{(3)}R^2}+\alpha_{1}\frac{1}{^{(3)}R} + \alpha_{2}~ ^{(3)}R^2+ \alpha_{3}~ ^{(3)}R^3,
\end{eqnarray}
where $\alpha_{i}$ $(i=0,..,3)$, are coupling constants. While the terms connected to $\alpha_{2}$ and $\alpha_{3}$ are due
to $UV$ corrections -- they might emerge from typical Ho\v{r}ava-Lifshitz potentials -- the terms linked to
$\alpha_{0}$ and $\alpha_{1}$ are objects of extreme large scale corrections where the diffeomorphism covariance may be broken as well. We will show that in this framework it is possible
to construct a concrete bouncing model with a late-time accelerated phase due to the spatial curvature term connected to $\alpha_1$.
The lower order term connected to $\alpha_{0}$ is introduced in order to probe the
final fate of the universe.

\section{A Concrete Model From Observational Constraints}

By defining the usual density parameters
\begin{eqnarray}
\Omega_{0m}=\frac{\rho_{0m}}{\rho^c_{0}},~~\Omega_{0r}=\frac{\rho_{0r}}{\rho^c_{0}},~~\Omega_{0k}=\frac{k}{a_0^2H^2_{0}},
\end{eqnarray}
where $\rho_0^c \equiv 3H_0^2/\kappa^2$,
the Hamiltonian constraint (\ref{Hamilton})-(\ref{Hamilton1}), can be rewritten in the form
\begin{eqnarray}
\label{eq4}
{\cal H}=-\frac{p_a^2}{24a}+{V}(a)=0
\end{eqnarray}
where
\begin{eqnarray}
\label{eq5}
{V}(a)=12a^3\times\frac{H^2_0}{2}\Big(\frac{\Omega_{0m}}{a^3}+\frac{\Omega_{0r}}{a^4}-\frac{\Omega_{0k}}{a^2}+\frac{ A_0 a^4}{216 \Omega_{0k}^2}
-\frac{A_1 a^2}{36  \Omega_{0k}}
+\frac{6A_2\Omega_{0k}^2}{a^4}-\frac{36A_3\Omega_{0k}^3}{a^6}\Big).
\end{eqnarray}
The coefficients $A_i$ are the respective coupling constants $\alpha_i$ rescaled:
\begin{eqnarray}
\label{eq5new}
A_0=\frac{\alpha_0}{H_0^6},~A_1=\frac{\alpha_1}{H_0^4},~A_2=\alpha_2 H_0^2,~A_3=\alpha_3 H_0^4.
\end{eqnarray}
It is worth noticing that the term connected to $A_2$ (or $\alpha_2$) behaves like a radiation component.

From (\ref{eq4}) we derive the dynamical system
\begin{eqnarray}
\label{eqm}
\dot{a}=\frac{\partial {\cal H}}{\partial p_a}=-\frac{p_a}{12a},~~~~~~~~
\dot{p}_a=-\frac{\partial {\cal H}}{\partial a}=-\frac{p_a^2}{24a^2}-\frac{d{ V}}{da}
\end{eqnarray}
so that $(a, p_a)$ are canonical variables. Given the above equations
we see that the structure of the phase space may be organized by critical points
$(a, p_a)=(a_i, 0)$, where $a_i$ are connected to the extrema of the potential $V(a)$.
In fact, as we will see in the following, given the observational parameters,
the phase space dynamics allow at least three critical points: two centers
separated by a saddle.

In order to compare
our hamiltonian constraint to the first Friedmann equation
for the $\Lambda$CDM model, it is useful to rewrite (\ref{eq4})-(\ref{eq5}) as
\begin{eqnarray}
\label{eqf}
\frac{1}{2}\Big(\frac{p_a}{12a}\Big)^2+{\cal{V}}(a)=0
\end{eqnarray}
where
\begin{eqnarray}
\label{eqf1}
{\cal V}(a)=\frac{H^2_0a^2}{2}\Big(\frac{\Omega_{0k}}{a^2}-\frac{ A_0 a^4}{216 \Omega_{0k}^2}
+\frac{A_1 a^2}{36  \Omega_{0k}}
-\frac{6A_2\Omega_{0k}^2}{a^4}+\frac{36A_3\Omega_{0k}^3}{a^6}-\frac{\Omega_{0m}}{a^3}-\frac{\Omega_{0r}}{a^4}\Big),
\end{eqnarray}
and $H\equiv \dot{a}/{a}$.

We now proceed to feed our model with observational parameters.
In (\ref{eqf1}) we see explicitly that all the corrections in our model emerge from the assumption of a nonvanishing
spatial curvature. As Planck data\cite{planck} still leave some room for curvature, namely,
\begin{eqnarray}
\label{ok}
\Omega_{0k}=0.001\pm 0.002,
\end{eqnarray}
the bounce condition $A_3 \Omega_{0k}^3>0$ may be satisfied for a positive or negative spatial curvature.

Taking into account the Planck data\cite{planck} for the matter and radiation density parameters,
we fix:
\begin{eqnarray}
\label{eq6}
\Omega_{0m}=0.315,~~
\Omega_{0r}=10^{-5}.
\end{eqnarray}
Furthermore, current observations\cite{planck,ioav} give support to the following constraints: (i)  ${\cal V}(a_0=1)=-H_0^2/2$; (ii) the deceleration parameter is given by $q_0 = -(a\ddot{a}/\dot{a}^2)|_0\simeq-0.54$; (iii) the predicted value of the scale factor $a_e$ -- at the end of the matter era -- corresponds to a redshift $z\simeq 0.4$ (or, $a_e \simeq 0.7$). The latter constraint implies that a
bouncing epoch followed by a decelerated phase with a graceful exit to a late-time accelerated regime can only be obtained as long as the potential ${\cal V}(a)$ has at least two local extrema -- one local minimum $a_1$ and
one local maximum $a_{e}$ -- so that $a_1<a_e$ and ${\cal V}(a_{e})<0$. Given (\ref{eq6}) it is easy to verify that
\begin{eqnarray}
(i)~{\cal V}(a_0=1)=-\frac{H_0^2}{2} \rightarrow A_1 \simeq \frac{0.16 A_0}{\Omega_{0k}} + \Omega_{0k}\{
-24.66 + \Omega_{0k} [-36 + \Omega_{0k}(216 A_2  - 1296 A_3 \Omega_{0k})]\};
\\
\nonumber
\\
(ii)~q_0 \simeq  -0.54 \rightarrow A_2 \simeq \frac{1}{\Omega_{0k}^4}\{0.00025 A_0 + \Omega_{0k}^2 [0.03736 + \Omega_{0k}(0.1  + 8 A_3 \Omega_{0k}^2)]\};~~~~~~~~~~~~~~
\\
\nonumber
\\
(iii)~\frac{d{\cal V}}{da}\Big|_{a_e}=0 \rightarrow 0.0064 A_0 + \Omega_{0k}^2 [0.6588 +
    \Omega_{0k}(1.71  - 280.22 A_3 \Omega_{0k}^2)]\simeq 0; ~~~~~~~~~~~~~~~~~~~
\\
\nonumber
\\
\nonumber
a_1<a_e~~{\rm and}~~{\cal V}(a_e)<0
    \rightarrow A0 \gtrsim \Omega_{0k}^2 (-284.94 -
    522.80 \Omega_{0k} +
    140136.01 A3 \Omega_{0k}^3).~~~~~~
\end{eqnarray}
From the above we see that the observations constrain our model to a certain region of the parametric space $(A_0, \Omega_{0k}, A_3)$. In Fig. 1 we plot such a domain generated by $(29)$, given (\ref{eq6})-$(28)$. In the left (right) panel we consider the case $\Omega_{0k}<0$ ($\Omega_{0k}>0$) so that the bounce condition reads $A_3<0$ ($A_{3}>0$). These plots were generated considering
the whole range (\ref{ok}) allowed by Planck data\cite{planck}. In both plots we see that $A_0<0$ so that a decelerated phase
is predicted after the late-time accelerated regime and the universe inexorably
recollapses. Therefore, the overall behaviour expected for our model is a universe in eternal recurrence.
%
%
%
\begin{figure}[tbp]
\centering
\begin{center}
\includegraphics[width=8.5cm,height=7.9cm]{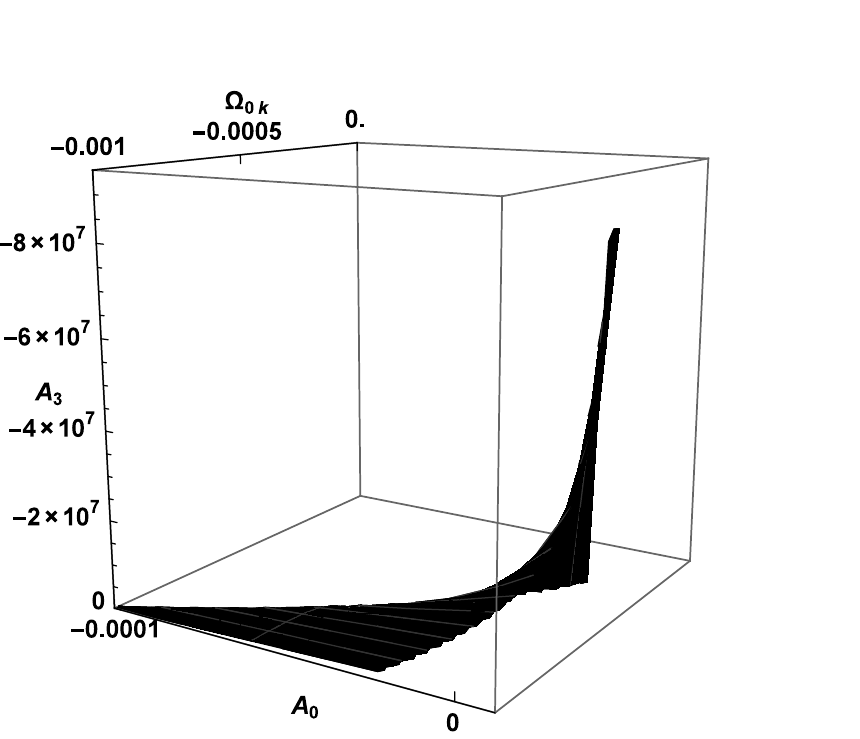}~\includegraphics[width=8.1cm,height=7.9cm]{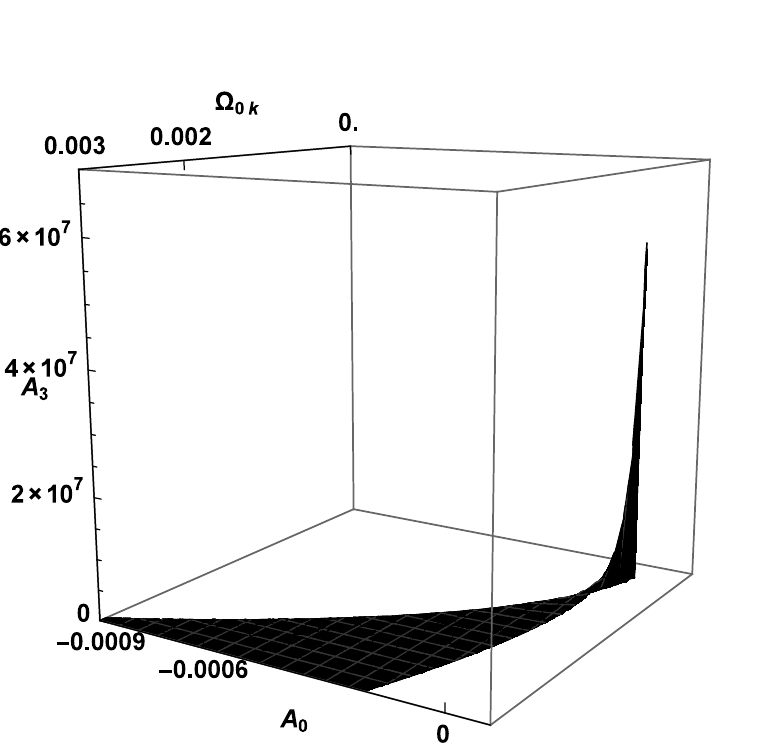}
\caption{\label{fig:i} The domain of $A_0$ -- as a function of $A_3$ and $\Omega_{0k}$ -- given the observational constraints (i), (ii) and (iii), see text. In both plots we see that $A_0<0$ so that a decelerated phase
is predicted after the late-time accelerated regime and the universe inexorably
recollapses. As we are considering only bouncing configurations, the overall behaviour of our model is a universe in eternal recurrence
within the range of $\Omega_{0k}$ allowed by Planck data.}
\end{center}
\end{figure}
%
\begin{figure}[tbp]
\centering
\begin{center}
\includegraphics[width=8.50cm,height=7.0cm]{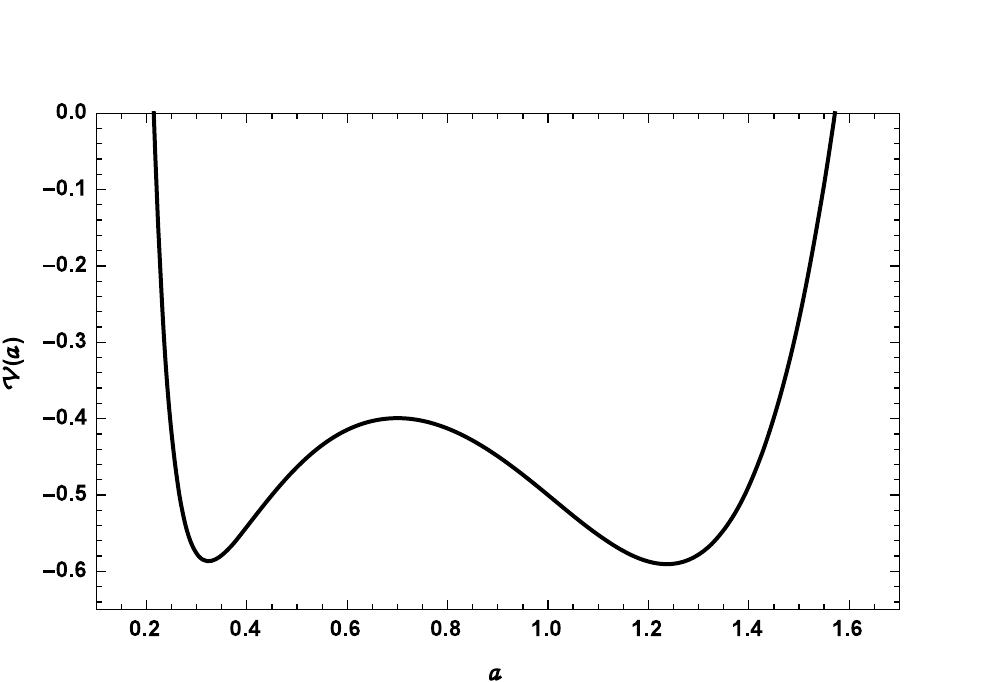}\includegraphics[width=8.50cm,height=6.5cm]{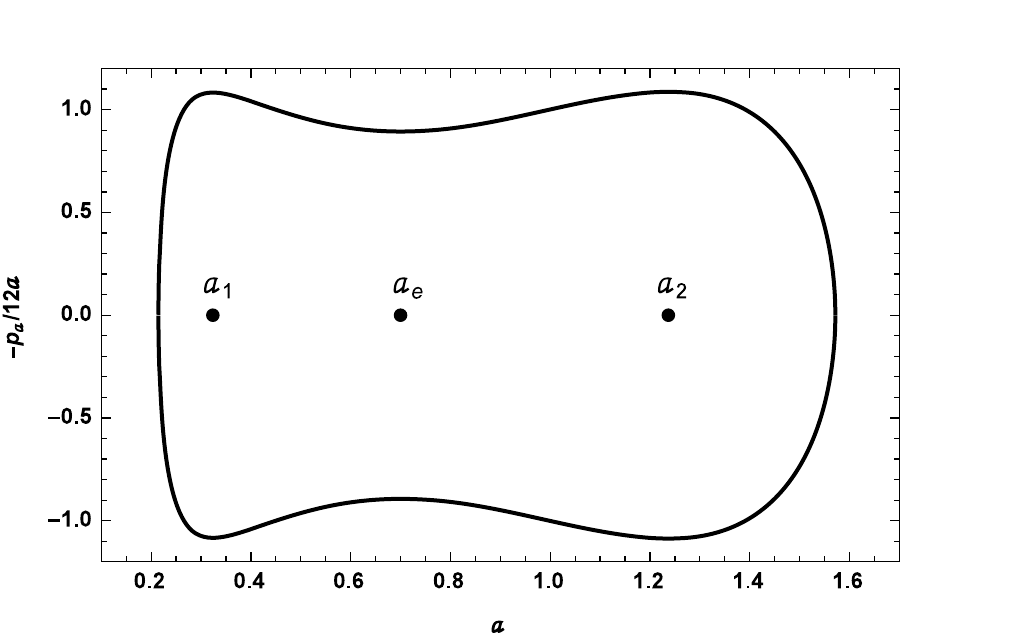}
\caption{\label{fig:ii}The behaviour of the potential ${\cal V}(a)$ (left panel) and the phase space (right panel). For the purpose of illustration, here we have fixed
$A_3=2\times 10^5$, $\Omega_{0k}=0.001$ and $H_0^2=1$. For $p_a>0$ we characterize three distinct regions: (i) the point $a=a_1$ sets up the transition from an early universe to a decelerated
radiation/matter era; (ii) $a=a_e$ defines the transition to a late-time accelerated regime; (iii) $a=a_2$ determines the end
of late-time acceleration and the universe starts its own recollapse towards an eternal recurrence configuration.}
\end{center}
\end{figure}
\begin{figure}[tbp]
\centering
\begin{center}
\includegraphics[width=8.50cm,height=7.5cm]{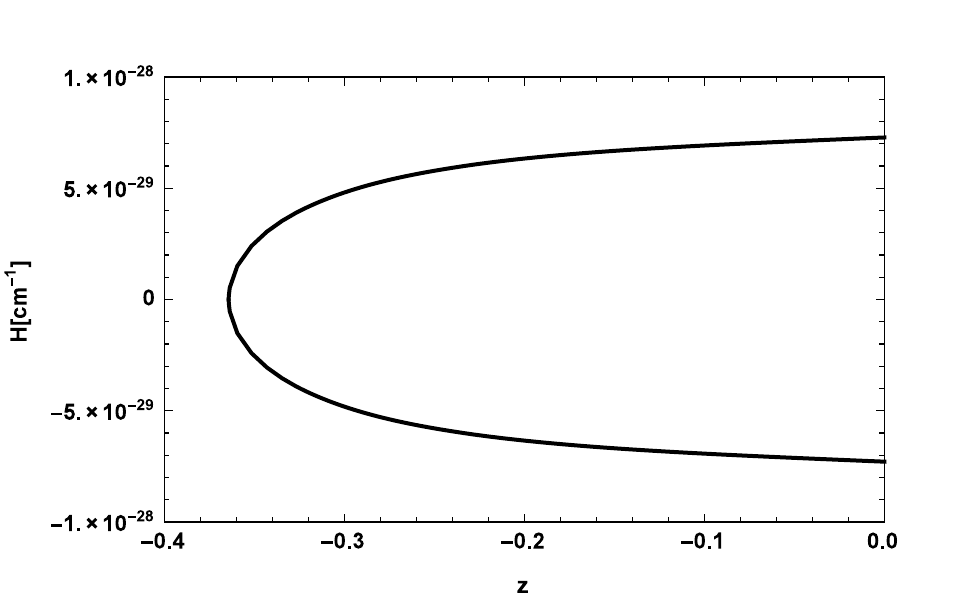}\includegraphics[width=8.50cm,height=7.4cm]{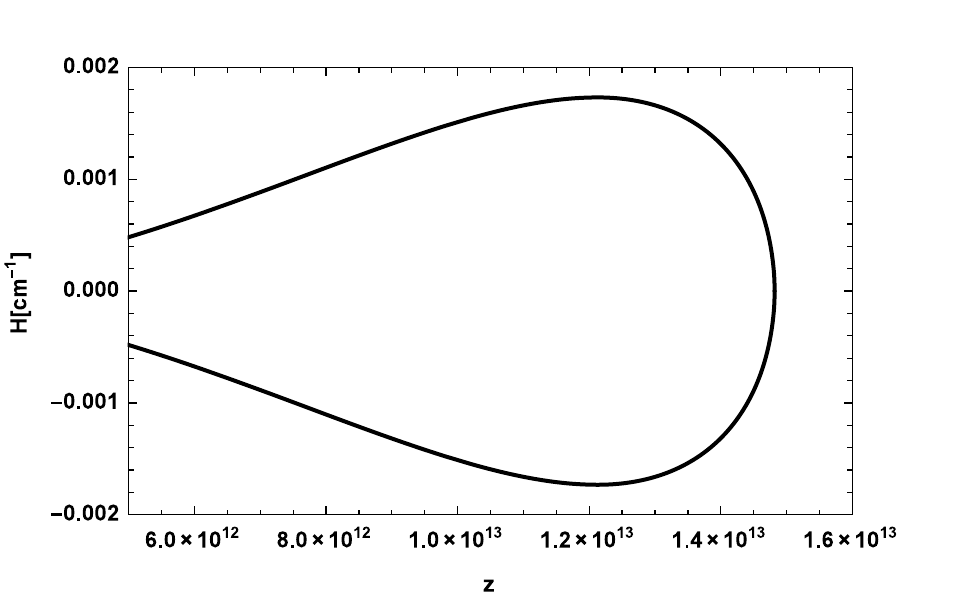}~~
\caption{\label{fig:iii} $H$ as a function of the redshift for $A_3=10^{-20}$. In the left panel
we show the behaviour of $H$ in a neighbourhood our present era until the universe recollapses at $z\simeq -0.36$. In the right panel
we show the behaviour of $H$ in a neighbourhood of the bounce.
Here we have fixed $H_0=7.28\times 10^{-29}{\rm cm}^{-1}$ in units $c=G=1$, so that $\alpha_0\simeq -7.6\times 10^{-503}{\rm GeV}^{-6}$,
$\alpha_1\simeq -3.2\times 10^{-334}{\rm GeV}^{-4}$, $\alpha_2\simeq 1.4\times 10^{170} {\rm GeV}^{2}$, $\alpha_3\simeq 1.1\times 10^{312} {\rm GeV}^{4}$.}
\end{center}
\end{figure}
%
As an illustration, we fix
$A_3=2\times 10^5$ and $\Omega_{0k}=0.001$.
In Fig. \ref{fig:ii} we show the behaviour of the potential ${\cal V}$ (left panel) and the phase space (right panel).

Now we proceed to put a better limit on the bounce parameter $A_3$ -- as expected, the smaller is $|A_3|$, smaller is the bounce scale. For that matter we are going to restrict ourselves to the case $\Omega_{0k}=0.001$.
Considering the evolution of quantum cosmological perturbations,
it has been shown\cite{npn} that in order to obtain amplitudes and wavelength spectra
compatible with CMB data, one must
satisfy the condition $l_c\gtrsim 10^3\times l_p$, where $l_c\propto 1/\sqrt{R}$ is the curvature scale at the bounce and $l_p$ is the Planck length. For that matter we infer the lower limit $A_3\gtrsim 10^{-52}$. On the other hand, if one intends to
not spoil predicted high-redshift events
like the cosmic neutrino background\cite{pastor}
-- at a redshift $z \simeq 10^{10}$ -- we fix the upper limit $A_3\lesssim 10^{-14}$.
Therefore, for every value of $A_3$ in the domain
\begin{eqnarray}
\label{eqa3}
10^{-52}\lesssim A_3\lesssim10^{-14},
\end{eqnarray}
we obtain a phase space orbit with the same shape of that one depicted in the right panel of Fig. 2. It is worth noticing that its shape is maintained for $38$ orders of magnitude. Given the physical range of interest (\ref{eqa3}) of $A_3$, from (\ref{eq6})-$(29)$ we obtain:
\begin{eqnarray}
A_0\simeq -0.0000944641,~~A_1\simeq -0.037594,~~A_2\simeq 13174.2.
\end{eqnarray}
As $A_0<0$ and $A_1<0$, we obtain a concrete cosmological model with a late-time accelerated regime in eternal recurrence, as expected. In order to better illustrate this behaviour, let us consider the Friedmann equation
as a function of the redshift $z$:
\begin{eqnarray}
\label{eqfriedmann}
H^2+{\cal U}(z)=0
\end{eqnarray}
where
\begin{eqnarray}
\label{eq51}
\nonumber
{\cal U}(z)=H^2_0\Big[\Omega_{0k}(1+z)^2-\frac{ A_0 }{216(1+z)^4 \Omega_{0k}^2}
+\frac{A_1 }{36(1+z)^2  \Omega_{0k}}
-6A_2(1+z)^4\Omega_{0k}^2\\
+36A_3(1+z)^6\Omega_{0k}^3-\Omega_{0m}(1+z)^3-\Omega_{0r}(1+z)^4\Big].~~~
\end{eqnarray}
In Fig. \ref{fig:iii} we show the behaviour of $H$ as a function of the redshift for $A_3=10^{-20}$. In this context, the domain of breaking diffeomorphism invariance -- where the model becomes invariant over foliations which preserve diffeomorphism -- is given by $-0.2\lesssim z $ and $10^{10}\lesssim z$. It is worth noting that the end of late-time acceleration should take place at a redshift $z\simeq -0.2$.

\section{Conclusion}

In this paper we construct a closed FLRW universe by imposing an anisotropic
scaling between space and time at extremely high and low energies.
From the Hamiltonian constraint we obtain a first integral
corresponding to a modified Friedmann equation
which contain additional terms arising from curvature.
Such terms implement nonsingular bounces in the early Universe
together with a late-time acceleration regime.
The matter content of the model is a nonrelativistic pressureless perfect fluid and radiation.
Considering the breaking of covariance diffeomorphism also at extreme large scales, we introduce a higher order term to probe the
final fate of the universe.
Given observational parameters we constrain the model so that an eternal recurrence
regime is inexorable. According to observational parameters the model predicts that
the end of late-time acceleration should take place at a redshift $z\simeq -0.2$. It is worth to remark that
this feature is maintained for a domain of $38$ orders of magnitude of the parametric space.

Our treatment in the paper is based strongly on the
Hamiltonian formulation, with a conserved Hamiltonian
constraint. By the use of canonical variables we
were able to make a global examination
of the phase space so that appropriate critical
points provide the bounce and a late-time acceleration regime.
Both epochs are due to the breaking of diffeomorphism covariance
which occurs at redshift $-0.2 \gtrsim z $ and $10^{10}  \lesssim z$,
respectively. In such domains the model becomes invariant over foliations which preserve diffeomorphism
instead.

The breaking of diffeomorphism invariance has been a topic of interest during
the last decade. In fact, several authors have argued
that such a feature is relevant for theories in which General Relativity is
an emergent phenomenon from a more
fundamental theory. By comparing our model to the Ho\v{r}ava-Lifshitz
scenario -- the most investigated framework in
which diffeomorphism invariance can be broken -- we would be in a position to
better understand the origin of the terms which provide the bounce together with
late-time acceleration in this paper.

In the framework of Ho\v{r}ava-Lifshitz,
although the projectable version seems to be plagued with
an extra scalar degree of freedom which is either classically unstable or
a ghost in the IR\cite{sotiriou}, there is still some room for a healthy
version of a proper theory of gravitation in which diffeomorphism covariance can be broken.
In fact, in the non-projectable version of Ho\v{r}ava-Lifshitz one may also include invariant contractions
of $\partial \ln{N} / \partial x^i$ in the potential $U$.  Connected to the lowest order invariant $\partial_i \ln{N} \partial^i\ln{N}$, there
is a coupling parameter $\sigma$ which defines a ``safe'' domain of the theory\cite{sotiriou,sotiriou3}.
Although in this
case there is also an extra scalar degree of freedom when one linearizes the theory in a
Minkowski background, for $0 < \sigma < 2$ this mode might not be a ghost
nor classically unstable (as long as detailed balance is not imposed).
Notwithstanding the non-projectable version also exhibits a strong coupling\cite{sotiriou3}-\cite{sotiriou4}, it has been argued that its scale is too high to be phenomenologically accessible from gravitational experiments\cite{sotiriou}.

Although the terms connected to the coupling constants $\alpha_2$ and $\alpha_3$ are genuine Ho\v{r}ava-Lifshitz potential corrections -- despite its versions -- terms such as those linked to $\alpha_0$ and $\alpha_1$ are not. In fact, the latter might
turn the Ho\v{r}ava-Lifshitz framework nonrenormalizable by power-counting.
As a future perspective we aim to better understand, from the theoretical point of view, what are the physical consequences/issues of such terms,
characterizing this depart from Ho\v{r}ava-Lifshitz gravity.


\begin{thebibliography}{99}

\bibitem{mukhanov} V. Mukhanov, {\it Physical Foundations of Cosmology} (Cambridge University
Press, 2005).
%
\bibitem{wald} R. M. Wald, {\it General Relativity} (University of Chicago Press, Chicago,
1984).
%
\bibitem{riess}
A. G. Riess et al., Astron. J. 116, 1009 (1998); S.
Perlmutter et al., Astrophys. J. 517, 565 (1999); D. Rubin et al., Astrophys. J. 695, 391
(2009); M. Hicken et al., Astrophys. J. 700, 1097
(2009).
%
\bibitem{ccp}  S. Weinberg, Rev. Mod. Phys. 61 1-23 (1989).
%
\bibitem{nelson0} Rodrigo Maier, Nelson Pinto-Neto, Ivano Dami\~ao Soares,
Phys. Rev. D 87, 043528 (2013).
%
\bibitem{nelson} Rodrigo Maier, Stella Pereira, Nelson Pinto-Neto, and Beatriz B. Siffert, Phys. Rev. D 85, 023508 (2012).
%
\bibitem{lta} Luca Amendola and Shinji Tsujikawa, {\it Dark Energy: Theory and Observations} (Cambridge University
Press, 2010).
%
\bibitem{dono} J. F. Donoghue, Phys. Rev. D 50, 3874 (1994).
%
\bibitem{rovelli} M. Gaul and C. Rovelli, ``Loop quantum gravity and
the meaning of diffeomorphism invariance'', Lect. Notes
Phys. 541, 277 (2000) [arXiv:gr-qc/9910079].
%
\bibitem{Horava} P. Ho\v{r}ava, Phys. Rev. D 79, 084008 (2009).
%
\bibitem{cal} G. Calcagni, JHEP 0909, 112 (2009).
%
\bibitem{brand} R. Brandenberger, Phys. Rev. D 80, 043516 (2009).
%
\bibitem{maier} Rodrigo Maier, Class. Quantum Grav. 30 (2013) 115011.
%
%
\bibitem{maier1} R. Maier and I. Dami\~ao Soares, Phys. Rev. D 96, 103532 (2017).
%
\bibitem{sotiriou2} C. Charmousis, G. Niz, A. Padilla and P.M. Saffin, JHEP 0908 070 (2009); D. Blas, O. Pujolas and S. Sibiryakov JHEP 0910:029 (2009); C. Bogdanos, Emmanuel N. Saridakis, Class. Quant. Grav. 27 075005 (2010); K. Koyama and F. Arroja, JHEP 1003 061 (2010); T. P. Sotiriou, M. Visser and S. Weinfurtner, JHEP 0910 033 (2009).
%
\bibitem{adm} R. Arnowitt, S. Deser and C. W. Misner, {\it Gravitation: an introduction to current research} (John Wiley \& Sons, New York, 1962, chapter 7).
%
\bibitem{cannone} Dario Cannone,  Gianmassimo Tasinato and David Wands, JCAP 1501 (2015) no.01, 029.
%
\bibitem{graef} L.L. Graef and R. Brandenberger, JCAP 10 (2015) 009.
%
\bibitem{planck} Planck 2018 results. VI. Cosmological parameters, arXiv:1807.06209.
%
\bibitem{ioav} M. Vargas dos Santos et al., JCAP02(2016)066.
%
\bibitem{npn} P. Peter, E. J. C. Pinho, and N. Pinto-Neto, Phys. Rev.
D 75 023516 (2007); Nelson Pinto-Neto, Physical Review D 79, 083514 (2009).
%
\bibitem{pastor} J. Lesgourgues and S. Pastor, Physics Reports. 429 (6): 307-379 (2006).
%
\bibitem{sotiriou} Thomas P. Sotiriou, J. Phys. Conf. Ser. 283 012034 (2011).
%
\bibitem{sotiriou3} T. P. Sotiriou, M. Visser and S. Weinfurtner, JHEP 0910 033 (2009).
%
\bibitem{niz} C. Charmousis, G. Niz, A. Padilla and P.M. Saffin, JHEP 0908 070 (2009).
%
\bibitem{sotiriou4} Antonios Papazoglou and Thomas P. Sotiriou, Phys. Lett. B 685 197-200 (2010).


\end{thebibliography}
\end{document}